\documentclass[aps,longbibliography,showpacs,twocolumn,superscriptaddress,amsmath,amssymb,verbatim]{revtex4-2}

\usepackage[LGR,T1]{fontenc}
\usepackage{mathptmx}
\usepackage[utf8]{inputenc}
\usepackage{color}
\usepackage{babel}
\usepackage{textcomp}
\usepackage{multirow}
\usepackage{graphicx}
\usepackage{amsmath}
\usepackage{soul}
\setstcolor{red}
\usepackage{mathtools}
\usepackage[unicode=true,pdfusetitle,
bookmarks=true,bookmarksnumbered=true,bookmarksopen=false,
breaklinks=true,pdfborder={0 0 0},pdfborderstyle={},backref=false,colorlinks=true]
{hyperref}
\hypersetup{
	linkcolor=blue, citecolor=blue,  urlcolor=blue}
\mathchardef\mhyphen="2D 
\def\tcb/{\textcolor{blue}}
\def\half/{$\frac{1}{2}$}
\def\yb3+/{Yb$^\mathrm{3+}$}
\def\jeff/{$J_{\mathrm{eff}}$}
\def\scto/{Sr$_3$CuTa$_2$O$_9$}
\def\ybsc/{Yb$_3$Sc$_2$Ga$_3$O$_{12}$}
\def\mnsn/{MnSnB$_2$O$_6$}
\def\mn2+/{Mn$^{2+}$}
\def\zncu/{ZnCu$_3$(OH)$_6$Cl$_{12}$}
\def\ybmg/{YbMgGaO$_4$}
\def\rucl/{$\alpha$-RuCl$_3$}
\def\musr/{$\mu$SR}

\DeclareUnicodeCharacter{2212}{-}

\begin{document}
	\preprint{APS/123-QED}
	\title{Possible realization of a randomness-driven quantum disordered  state in an $S=1/2$ antiferromagnet \scto/}
	
	\author{B. Sana}
	\affiliation{Department of Physics\unskip, Indian Institute of Technology Madras\unskip, Chennai\unskip, 600036, India\\ }
	
	\author{M. Barik}
	\affiliation{Department of Physics\unskip, Indian Institute of Technology Madras\unskip, Chennai\unskip, 600036, India\\ }
	\author{S. Lee}
	\affiliation{Center for Integrated Nanostructure Physics\unskip, Institute for Basic Science (IBS)\unskip, Suwon 16419, Republic of Korea\\ }
	\author{U. Jena}
	\affiliation{Department of Physics\unskip, Indian Institute of Technology Madras\unskip, Chennai\unskip, 600036, India\\ }
	
	\author{M. Baenitz}
	\affiliation{Max Planck Institute for Chemical Physics of Solids\unskip, Nöthnitzer Strasse 40\unskip, 01187 Dresden\unskip, Germany}
	\author{J. Sichelschmidt}
	\affiliation{Max Planck Institute for Chemical Physics of Solids\unskip, Nöthnitzer Strasse 40\unskip, 01187 Dresden\unskip, Germany}
	\author{S. Luther}
	\affiliation{Hochfeld-Magnetlabor Dresden (HLD-EMFL), Helmholtz-Zentrum Dresden-Rossendorf, 01328 Dresden, Germany}
	\affiliation{Institut f\"{u}r Festk\"{o}rper- und Materialphysik, TU Dresden, 01062 Dresden, Germany}

	\author{H.~K\"{u}hne}
	\affiliation{Hochfeld-Magnetlabor Dresden (HLD-EMFL), Helmholtz-Zentrum Dresden-Rossendorf, 01328 Dresden, Germany}

	\author{ K. Sethupathi}
	\affiliation{Department of Physics\unskip, Indian Institute of Technology Madras\unskip, Chennai\unskip, 600036, India\\ }
	\affiliation{Quantum Centre for Diamond and Emergent Materials\unskip, Indian Institute of Technology Madras, Chennai\unskip, 600036, India\\}
	\author{M. S. Ramachandra Rao}
	\affiliation{Quantum Centre for Diamond and Emergent Materials\unskip, Indian Institute of Technology Madras, Chennai\unskip, 600036, India\\}
	\affiliation{Department of Physics\unskip, Nano Functional Materials Technology Centre and Materials Science Research Centre\unskip, Indian Institute of Technology Madras,
	Chennai\unskip, Tamil Nadu 600036, India\\}
	
	\author{K. Y. Choi}
	\affiliation{Department of Physics\unskip, Sungkyunkwan University\unskip, Suwon 16419, Republic of Korea}
	\author{P. Khuntia}
	\email{pkhuntia@iitm.ac.in}
	\affiliation{Department of Physics\unskip, Indian Institute of Technology Madras\unskip, Chennai\unskip, 600036, India\\ }
	\affiliation{Quantum Centre for Diamond and Emergent Materials\unskip, Indian Institute of Technology Madras, Chennai\unskip, 600036, India\\}
	

\begin{abstract}
 Collective behavior of spins, frustration-induced strong quantum fluctuations and subtle interplay between competing degrees of freedom in quantum materials can lead to correlated  quantum states with exotic excitations that are essential ingredients for establishing paradigmatic models and have immense potential for quantum technologies. Disorder is ubiquitous in real materials, and the detailed insights into the role of disorder on the intriguing ground state borne out of quenched randomness provide a route towards the design and discovery of functional quantum materials. Herein, we report magnetization, specific heat, electron spin resonance, and muon spin resonance studies on a 3$d$-electron based  antiferromagnet \scto/. The negative value of Curie-Weiss temperature, obtained from the Curie-Weiss fit of high temperature magnetic susceptibility  data indicates the presence of antiferromagnetic interaction between Cu$^{2+}$ moments.  Specific heat data show the absence of long-range magnetic ordering down to 64 mK  despite a reasonably strong exchange interaction between Cu$^{2+}$ ($S=1/2$) spins as reflected from a Curie-Weiss temperature of $-27\pm1$ K. The power-law behavior and the data collapse of specific heat and magnetization data evince the emergence of a random-singlet state in \scto/. The power-law-like spin auto-correlation function and the data collapse of muon polarization asymmetry with longitudinal field dependence of $t/{(\mu_0H)}^{\gamma}$ further support credence to the presence of a randomness-induced quantum disordered  state. Our results suggest that randomness induced by disorder  is an alternate route to realize quantum spin disordered  state in this antiferromagnet.
\end{abstract}

\maketitle
\section*{I. Introduction}

Spin correlations, competing interactions and quantum character of  spins in frustrated quantum materials  offer an ideal platform to harbor entangled quantum states with fractional quantum numbers. These states could have significant implications in addressing some of the fundamental challenges in condensed matter and  the development of quantum technologies \cite{Balents2010,Savary_2016,RevModPhys.89.025003,RevModPhys.80.1083}. Magnetic frustration, which refers to the incompatibility of various magnetic interactions in the spin-lattice of a quantum material, can lead to a highly degenerate ground state and prevent symmetry-breaking phase transitions in quantum materials.  Frustrated magnets are ideal to host novel quantum states such as quantum spin liquids and spin ice \cite{Balents2010,Ramirez1999}. A quantum spin liquid (QSL) is a highly entangled state of matter wherein frustration-induced strong quantum fluctuations prevent long-range magnetic order down to absolute zero temperature despite strong exchange interactions between magnetic moments. Remarkably, QSL is characterized by  the presence of fractional excitations  such as spinons and Majorana fermions coupled to emergent gauge fields and the spin-correlations display  non-local behavior that is robust against  weak external  perturbations \cite{Balents2010,ANDERSON1973153,Savary_2016,RevModPhys.89.025003,Khuntia2020,PhysRevLett.116.107203,KHUNTIA2019165435,Gao2019}. The experimental realization of QSLs with fractional quasiparticles  in one-dimensional (1D) quantum magnets  owing to strong quantum fluctuations is a  well-established scenario. However, the realization of QSLs in two-dimensional (2D) and three-dimensional (3D) quantum magnets remains challenging in view of the presence of various perturbations such as disorder, defects, extra connectivity of magnetic moments in the spin-lattice and finite inter-layer interactions in real materials.  In this context, strong quantum fluctuations in frustrated magnets, next-nearest interaction, ring exchange, and exchange anisotropy offer a viable platform in hosting QSL states in promising 2D and 3D spin-lattices \cite{Balents2010,Arh2022}. 

The experimental realization of QSLs in pristine and periodic systems remain a great challenge that invokes to discover, design and investigate promising frustrated magnets in this context . Interestingly, it has been suggested that unavoidable  disorder in certain frustrated magnets, can lead to a random exchange energy between spins \cite{PhysRevB.102.094407,PhysRevX.8.041040,Kitagawa2018,PhysRevLett.124.047204,Kimchi2018,Khatua2022}. Quenched disorder owing  to the presence of defects or site mixing that are fixed into the crystal lattice can induce randomness in exchange  energies and hence quantum fluctuation, which offers an alternate route to stabilize exotic quantum states in quantum magnets \cite{PhysRevX.8.031028,Khatua2022}.  The subtle interplay between quantum fluctuations induced by magnetic frustration and competition between emergent degrees of freedom can give rise to novel disordered ground states such as random-singlet state, where spins interact antiferromagnetically and form spin-singlet pairs. In 1D quantum spin chain with Heisenberg antiferromagnetic interactions, random coupling between $S=1/2$ spins forms random spin-singlet pairs \cite{PhysRevLett.43.1434,PhysRevB.50.3799,PhysRevB.22.1305}. However, random-singlet states in 2D and 3D spin-lattices have not been well studied yet due to the scarcity of  such quantum magnets \cite{PhysRevB.22.1305,PhysRevB.50.3799,doi:10.1063/1.329684,PhysRevLett.48.344,PhysRevB.98.134427}.

Disorder-induced quantum spin liquid-like state due to the substitution or site sharing of nonmagnetic elements with magnetic ones is also proposed in a few promising frustrated materials such as Ho$_2$Ti$_2$O$_7$, Pr$_2$Zr$_2$O$_7$ \cite{PhysRevLett.118.087203,PhysRevLett.118.107206}, Ba$_3$CuSb$_2$O$_9$ \cite{PhysRevLett.115.147202} and Y$_2$CuTiO$_6$ \cite{PhysRevLett.125.117206}. It is proposed that disorder can induce long-range entanglement and depending upon the degree of disorder, it can show Coulombic spin liquid, Mott glass and conventional glassy phases in the ground state \cite{PhysRevLett.118.087203}. For instance, in  Y$_2$CuTiO$_6$, $S=1/2$ Cu$^{2+}$ ions and non-magnetic Ti$^{4+}$ ions  share the same atomic site forming a triangular lattice with a 50:50 occupancy ratio. Despite such a huge site dilution, Y$_2$CuTiO$_6$ does not show conventional magnetic order or spin freezing down to 50 mK, indicating a disorder-driven cooperative paramagnetic state. Interestingly, the universal scaling of specific heat in Y$_2$CuTiO$_6$ also reveals the formation of a random-singlet
state  at low temperatures
\cite{PhysRevLett.125.117206}. In Ba$_3$CuSb$_2$O$_9$, the CuSbO$_9$ unit forms Cu-Sb dumbbells, which give rise to electric dipole moment due to the charge difference between Cu$^{2+}$  and Sb$^{5+}$ ions, and the site sharing  of Cu$^{2+}$ (50\%) with non-magnetic Sb$^{5+}$ (50\%) ions leads to spin-orbital liquid-like behavior \cite{Nakatsuji559,PhysRevLett.115.147202}. In a similar vein, triangular lattice antiferromagnet Sr$_3$CuSb$_2$O$_9$ with substantial anti-site disorder between Cu$^{2+}$ (33\%) and Sb$^{5+}$ (67\%) ions shows quantum spin liquid-like behavior down to 65 mK \cite{PhysRevLett.125.267202}. Recently, a quantum spin-liquid-like random-singlet state is observed in disordered antiferromagnet Li$_4$CuTeO$_6$  and an organic verdazyl-based complex with a honeycomb lattice \cite{Khatua2022,Yamaguchi2017}. The substitution of suitable chemical element in a controlled manner  can reveal interesting insights in the host lattice of frustrated magnet, for instance, the substitution non-magnetic Ir$^{3+}$ ion  at Ru$^{3+}$ site of  the celebrated Kitaev magnet $\alpha$-Ru$_{1-x}$Ir$_x$Cl$_3$  induces quenched disorder that drives  the diluted system into a spin liquid state \cite{PhysRevB.102.094407}.

Due to bond randomness, there exists a broad range of antiferromagnetic exchange interactions that lead to a power-law distribution of exchange interactions $P(J)\sim J^{-\alpha}$ at low energies in the renormalization group flow of the random-singlet state \cite{PhysRevX.8.031028}. This is manifested as unconventional  scaling behavior of low-temperature magnetic susceptibility and
specific heat in frustrated magnets with quenched disorder \cite{PhysRevX.8.031028,doi:10.7566/JPSJ.86.044704,PhysRevLett.124.047204,Khatua2022}. Such a scenario is related to the random distribution of exchange couplings in the host spin-lattice owing to anti-site disorder, which might lead to the power law behavior of thermodynamic and microscopic observables evincing a gapless excitation spectra.  So far, we have discussed the materials wherein randomness is originating from the site sharing of nonmagnetic elements. Recently, the rare-earth based triangular lattice antiferromagnet YbMgGaO$_4$ has received widespread acclaim as a promising quantum spin liquid candidate \cite{Li2015,PhysRevLett.115.167203,Shen2016,Paddison2017}. The presence of anti-site disorder between Mg$^{2+}$ and Ga$^{3+}$ leads to the formation of a random crystal field, which is believed to induce bond disorder and give rise to  the emergence of quantum disordered liquid-like state \cite{PhysRevLett.119.157201}. A random-singlet state is demonstrated in a chemically substituted $S=1/2$ square lattice antiferromagnet Sr$_2$CuTe$_{0.5}$W$_{0.5}$O$_6$ \cite{Mustonen2018}, wherein the square lattice of Cu$^{2+}$ is not disordered but bond randomness originates from the site mixing of non-magnetic Te$^{6+}$ and W$^{6+}$ ions.  The spin-$1/2$ Heisenberg $J_1\mhyphen J_2$ model on a honeycomb lattice is also proposed to host random-singlet-driven spin-liquid-like behavior \cite{doi:10.7566/JPSJ.86.044704}. The nearest-neighbor antiferromagnetic Heisenberg interaction $J_1$ can not induce frustration in the bipartite honeycomb lattice and the ground state is antiferromagnetic . An additional next nearest-neighbor antiferromagnetic exchange interaction $J_2$ is essential to induce frustration that leads to a quantum spin-liquid-like ground state in a honeycomb lattice. The random distribution of exchange interactions leads to
 a wide range of binding energies for singlet dimers, resulting
 in a gapless behavior \cite{PhysRevX.8.041040}. 

The random-singlet state in 1D quantum spin systems has been well-established through extensive studies \cite{PhysRevB.99.035116,PhysRevLett.43.1434,PhysRevB.50.3799,PhysRevB.22.1305}. Even an infinitesimal bond disorder can lead a 1D spin system into a random-singlet state \cite{PhysRevB.61.11552,PhysRevLett.91.229701}. For instance,
substituting Si with Ge, a random-singlet state is achieved in the spin-chain  BaCu$_2$(Si$_{1-x}$Ge$_x$)$_2$O$_7$ \cite{PhysRevB.99.035116}, where the parent compound BaCu$_2$Si$_2$O$_7$ shows long-range antiferromagnetic order \cite{PhysRevB.88.054422}. Understanding the role of randomness in destabilizing N\'eel order and driving  the spin system towards  a quantum disordered state in 2D and 3D spin-lattice is an uncharted venue. How the random-singlet in 2D and 3D spin-lattices  resembles or differs from that found in 1D spin-lattices is still unclear. More experimental candidates and evidence is required to explore how randomness and/or frustration induces intriguing quantum states  in 2D and 3D spin-lattices  \cite{PhysRevX.8.041040,PhysRevB.98.134427}. Disorder is inevitable in real materials; in this context, the search and investigation of  magnetic materials on 2D and 3D spin lattices with quenched disorder leading to randomness-induced correlated quantum states has drawn significant attention recently \cite{Kimchi2018,PhysRevX.8.031028,PhysRevB.87.214417}.

 In this work, we report magnetization, specific heat, muon spin resonance, and electron spin resonance studies on Sr$_3$CuTa$_2$O$_9$ (henceforth SCTO) wherein magnetic Cu$^{2+}$ ions constitute a 3D nearly cubic  spin-lattice. In this material, Cu$^{2+}$ ion shares the same crystallographic site with non-magnetic Ta$^{5+}$ ion with a ratio $1:2$.  The magnetic susceptibility shows no signature of a magnetic phase transition and spin-freezing down to 2 K. The  Cu$^{2+}$ ($S=1/2$) moments interact antiferromagnetically with a Curie-Weiss temperature ($-27\pm1$ K) reflecting moderate exchange interaction between Cu$^{2+}$ moments. Specific heat confirms the absence of long-range magnetic ordering down to 64 mK, implying a disordered ground state.  The magnetization, specific heat and muon asymmetry  results display a data collapse behavior pointing towards  correlated quantum state.
 \begin{figure*}[!ht]
 	\includegraphics[width=16cm, height=7cm]{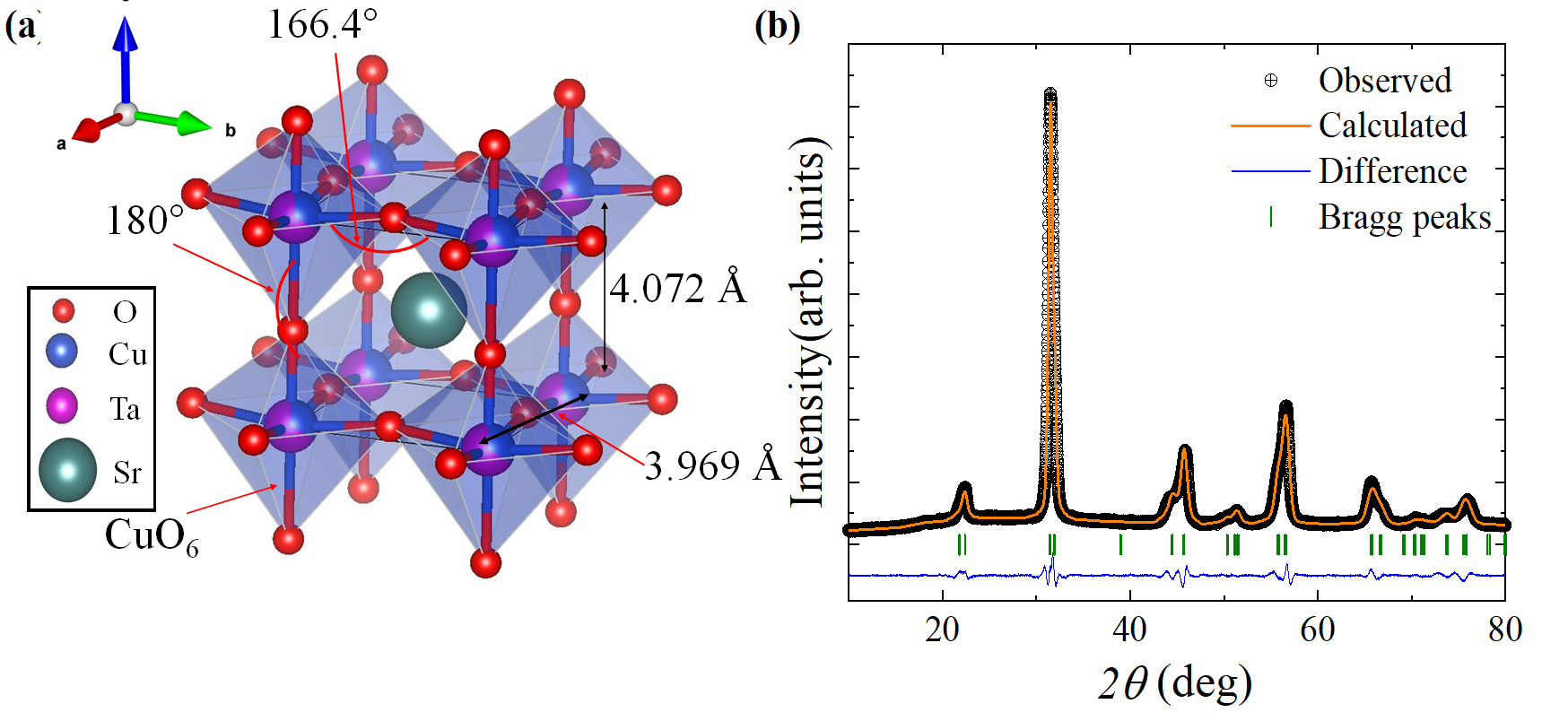}
 	\caption{\textbf{(a)} Crystal structure of \scto/.    \textbf{(b)} Rietveld refinement of powder XRD data recorded at room temperature.}{\label{Structure_SCTO}}
 \end{figure*}The unusual scaling behavior of high field magnetization up to 52 T, ESR line width and zero-field muon relaxation rate at low temperature along with the data collapse behavior of thermodynamic and $\mu$SR results suggest a random-singlet state in SCTO that is reconciled with the theoretical framework in the context of a randomness-induced quantum  disordered state.

\section*{II. METHODS}

Polycrystalline sample of SCTO was prepared by a conventional solid-state synthesis method from high purity Sr$_2$CO$_3$ (Alfa Aesar, 99.994\%), Ta$_2$O$_5$ (Alfa Aesar, 99.993\%) and CuO (Alfa Aesar, 99.995\%). The stoichiometric amount of each material was mixed together and  then pelletized.  The   single phase was obtained after firing it at 800 $^\circ$C (12 hrs), 900 $^\circ$C (24 hrs) and 950$^\circ$C (48 hrs) with intermediate grindings. The phase purity was checked by a Rigaku X-Ray diffractometer at room temperature using Cu $K_{\alpha}$ radiation. The physics of correlated quantum materials is quite rich and diverse, which requires complementary  experimental techniques to probe various facets at play. A SQUID-VSM instrument (Quantum Design)  was used to perform  magnetization measurements in the temperature range of 2 K $\leq T \leq$ 400 K under magnetic fields 0 T $\leq \mu_\mathrm{0}H \leq$ 5 T. Specific heat measurements were carried  out using a Physical Properties Measurement System (QD, PPMS) in the temperature range of  2 K $\leq T \leq$ 200 K and in magnetic fields of 0 T $\leq \mu_\mathrm{0}H \leq$ 7 T. In addition, low-temperature specific heat measurements in zero field and 1 T  were carried out down to 64 mK using a dilution refrigerator (Quantum Design).  We used a continuous wave ESR spectrometer at X-band frequencies ($\nu = 9.4 $ GHz). The  powder sample was embedded in
paraffine and the temperature was set in between 3 K $\leq T\leq$ 300 K using a helium-flow cryostat. We performed zero-field (ZF) and longitudinal-field (LF) muon spin relaxation (\musr/) measurements on the M20 beamline using the LAMPF spectrometer (2 K $\leq T\leq$ 200 K) at TRIUMF (Vancouver, Canada). The obtained $\mu$SR data were analyzed using MUSRFIT software \cite{SUTER201269}. The high-field magnetization experiment was conducted at 1.4 K in  pulsed magnetic fields up to 52 T using the facilities at the Dresden High Magnetic Field Laboratory.
\begin{figure*}[!htb]
	\centering
	\includegraphics[height=13cm,width=16cm]{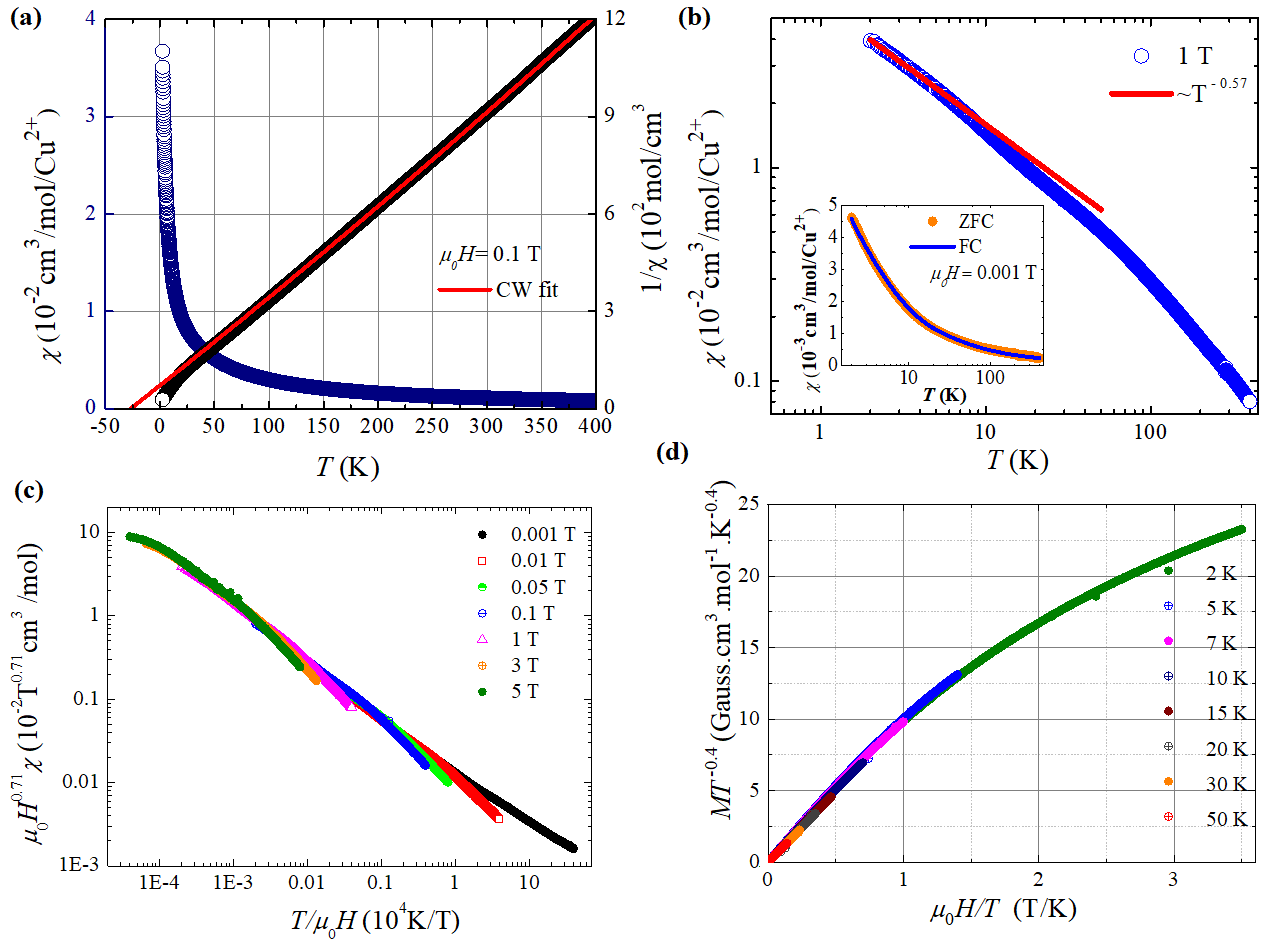}
	\caption{\textbf{ (a)} (Left $y$-axis): Temperature dependence of magnetic susceptibility of SCTO measured in an applied magnetic field of 0.1 T. (Right $y$-axis): The temperature dependence of inverse susceptibility data with Curie-Weiss fit.  \textbf{ (b)} The power-law fit of magnetic susceptibility in 1 T. The inset shows the ZFC-FC data recorded in 0.001 T. \textbf{ (c)} Scaling of magnetic susceptibility with $(\mu_\mathrm{0}H)^{0.71}\chi$ vs. $T/\mu_\mathrm{0}H$. \textbf{(d)} Scaling of magnetization isotherms with  $MT^{-0.4}$ vs. $\mu_0H/T$. The collapse of magnetic susceptibility and magnetization isotherms to a single curves suggests a randomness-induced quantum effect is at play in this antiferromagnet.}
	\label{sctomag}
\end{figure*}
 
\section*{III. Results}

\subsection*{Crystal structure}
Rietveld refinement of powder x-ray diffraction data was carried out using GSAS software \cite{Toby:hw0089}. For Rietveld refinement, initial structural parameters were taken from the isostructural compound Sr$_3$CuNb$_2$O$_9$ \cite{ZACHARIASZ2015627}. This compound is closely related to the perovskite family with general formula $AB$O$_3$, wherein the $A$ site is occupied by Sr and $B$ site is shared by Cu and Ta. \scto/ crystallizes in the tetragonal space group $P4mm$ \cite{Ono1992,zachariasz2019examination} (see Fig.\ref{Structure_SCTO}a). The Rietveld refinement parameters of SCTO are given in Table \ref{tab1} and the refinement pattern is shown in  Fig. \ref{Structure_SCTO}b.
\begin{table}[!ht]
	\centering
	\renewcommand{\arraystretch}{1.5}
	\renewcommand{\tabcolsep}{0.2cm}
	
	\begin{tabular}{|l|l|l|l|l|l|} 
		
		\hline
		Atoms & Wyc. pos.&x & y & z &Occupancy \\
		\hline
		Sr & 1$b$ &0.5 &0.5 &0.502(4)&1\\
		Cu & 1$a$ &0 &0 &0.014 &0.333 	
		\\
		Ta &1$a$ &0 &0 &0.014 &0.667
		
		\\
		O &1$a$ &0 &0 &0.471(5)
		
		&1
		\\
		O &2$c$ &0.5 &0 &0.068(6)&1
		
		\\
		
		\hline
		
	\end{tabular}
	\caption{Structural parameters for Sr$_3$CuTa$_2$O$_9$ determined from the Rietveld refinement of powder X-ray diffraction pattern using GSAS software. Space group $P4mm$ and cell parameters $a = b =$ 3.969(4) {\r{A}}, $c =$ 4.072(5)  {\r{A}}, $\alpha$ = $\beta$ = $\gamma$ = 90$^\circ$, $R_\mathrm{wp}=4.3\%$, $R_\mathrm{p}=3.2\%$, and $\chi^2 =4.8$.}{\label{tab1}}
\end{table}  The intra-layer Cu$^{2+}$-Cu$^{2+}$ bond length is 3.969(4) {\r{A}} whereas, the inter-layer length is 4.072(5) {\r{A}} (see Fig.\ref{Structure_SCTO}a). The nearest-neighbor Cu$^{2+}$ ions are connected via a Cu-O-Cu super-exchange path where the Cu-O-Cu angle is 166.4$^\circ$. The inter-planar super-exchange interaction is mediated via another Cu-O-Cu path with an angle of 180$^\circ$. This is in accordance with the Goodenough-Kanamori rule \cite{GOODENOUGH1958287} which suggests antiferromagnetic interaction is expected from the Cu-O-Cu exchange path. The comparable intra-layer and inter-layer distances makes  it a nearly cubic spin-lattice, but $2/3$ of the spin-lattice are shared by non-magnetic Ta$^{5+}$ ions. The Cu$^{2+}$ site occupancy in the spin lattice is $1/3$ which is close to the site  percolation threshold for a simple cubic lattice $p_\mathrm{c}=0.3116$ \cite{10.1093/acprof:oso/9780198570752.001.0001,PhysRevE.87.052107}. Six oxygen atoms constitute a CuO$_6$ polyhedron where Cu$^{2+}$ ion is at the center of the polyhedron (see Fig.\ref{Structure_SCTO}a). In this material, there is an unavoidable anti-site disorder between Cu$^{2+}$ and Ta$^{5+}$ that is close to the limit of the percolation threshold, and Cu$^{2+}$ $(S = 1/2$) spins constitute a nearly cubic spin-lattice with comparable interplane and intraplane bond distances.  It is interesting to investigate the role of quenched disorder in determining the ground state properties in this site-diluted three-dimensional antiferromagnet to understand the role of disorder in stabilizing intriguing quantum states in higher-dimensional spin-lattices.

\subsection*{Magnetization}

\begin{figure*}[!htb]
	\includegraphics[width=16cm, height=7cm]{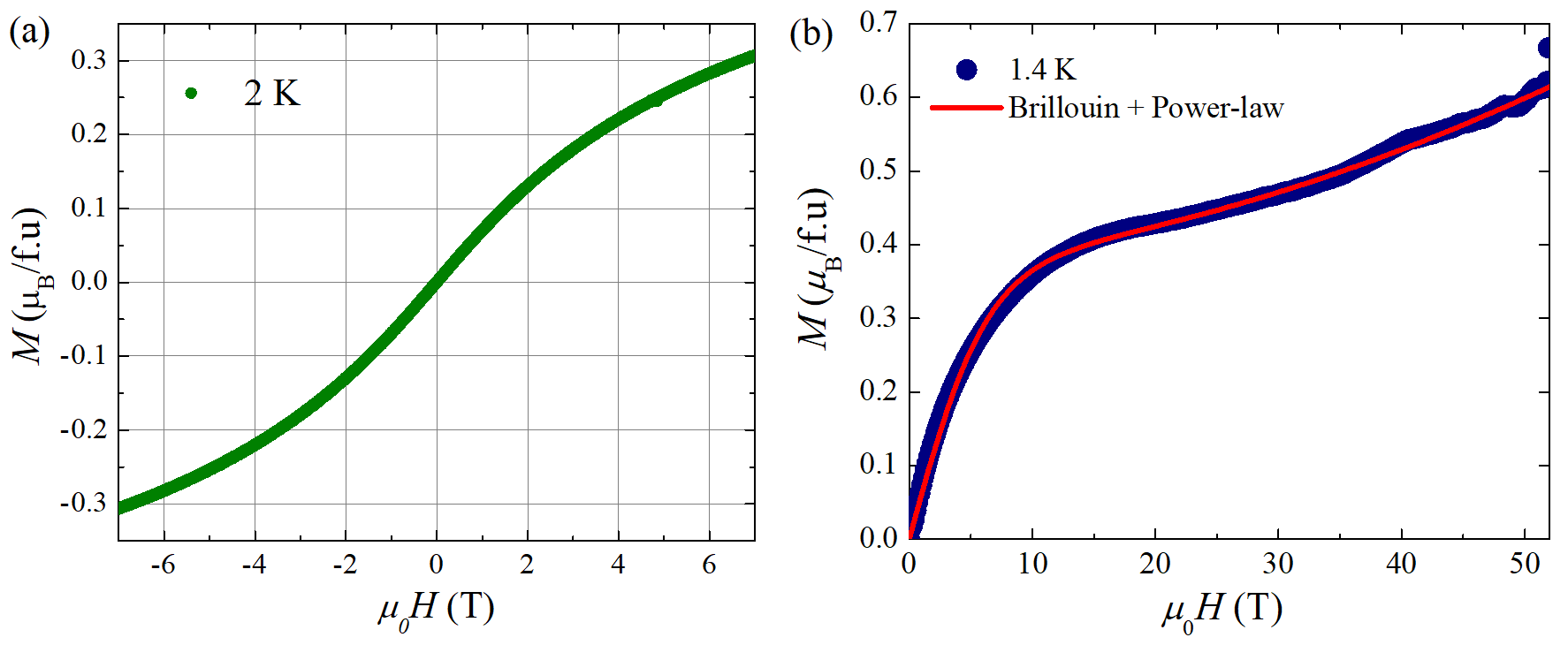}
	\caption{\textbf{ (a)} The absence of hysteresis in  $M$ vs. $\mu_0H$ curve taken at 2 K indicates that there is no impurity and no history dependence of magnetization despite huge site dilution. \textbf{(b)}  High-field magnetization isotherm at 1.4 K is denoted by filled circles. Solid red line indicates the Brillouin and power-law fit as described in the eqn.~\ref{MHBrillouin}. }{\label{MH_high}}
\end{figure*}
Magnetization experiments as a function of temperature and field   convey valuable information concerning  the nature of exchange interaction between magnetic moments in the spin-lattice, spin correlations, anisotropy, and disorder that are essential to shed insights into the ground state properties of quantum magnets. The left $y$-axis of Fig. \ref{sctomag}a represents the magnetic susceptibility taken in an applied magnetic field of 0.1 T in the temperature range 2 K $\leq T\leq$ 400 K. There is no anomaly in the magnetic susceptibility data down to 2 K, suggesting the absence of a magnetic phase transition. The inverse susceptibility data (see right $y$-axis of Fig. \ref{sctomag}a) can be well reproduced by the Curie-Weiss formula, $\chi(T)= \chi_0 +\frac{C}{T-\theta_\mathrm{CW}}$ in the temperature range 250 K $\leq T\leq$ 400 K. The value of the temperature independent susceptibility $\chi_0$ is  $-6.2\times$10$^{-5}$ cm$^3$/mol, which is originated due to core diamagnetic susceptibility as well as van-Vleck paramagnetic susceptibility. However, the negative sign indicates that diamagnetic susceptibility dominates the latter. A negative Curie-Weiss temperature of  $\theta_\mathrm{CW}\approx - 27\pm1$ K was extracted from the high-temperature Curie-Weiss fit, which indicates that the dominant interaction between the Cu$^{2+}$ spins is antiferromagnetic. The Curie-Weiss fit yields an effective moment of $\mu_{\mathrm{eff}} = 1.74\pm0.03$ $\mu_\mathrm{B}$, which is close to the free ion moment of $S=1/2$ Cu$^{2+}$ ion ($g\sqrt{S(S+1)}\mu_\mathrm{B}$, $g=$ Land\'e $g$ factor). The zero-field-cooled (ZFC) and field-cooled (FC) data taken in an applied magnetic field of 0.001 T (see inset of Fig. \ref{sctomag}b) do not show any bifurcation, suggesting the absence of any spin freezing. In spite of huge anti-site disorder between Cu$^{2+}$ and Ta$^{5+}$, the Cu$^{2+}$  site occupation seems correlated, keeping the magnetic connectivity intact with a moderate antiferromagnetic exchange interaction reflected as a Curie-Weiss temperature of $-27\pm1$ K.

 In quenched-disorder materials, random exchange interaction gives rise to a power-law behavior ($\propto T^{-\gamma}$) in low-temperature magnetic susceptibility data \cite{PhysRevX.8.031028,Kimchi2018,PhysRevB.102.054443}. We fitted the  susceptibility data  (see Fig. \ref{sctomag}b) in between 2-5 K with a power-law $\chi(T)\propto T^{-0.57}$. Another key signature of the random-singlet state is the scaling of susceptibility and specific heat data \cite{PhysRevLett.124.047204}. We plotted $(\mu_\mathrm{0}H)^{\gamma}\chi$ vs. $T/\mu_0H$ in Fig.~\ref{sctomag}c for magnetic field up to 5 T where the value $\gamma$ turned out to be 0.71, which is close to the value obtained from the power-law fit. All data points fall into the same curve at low temperatures, indicating bond randomness in the system \cite{PhysRevLett.122.167202}.  To confirm the universal scaling behavior, $MT^{-\beta}$ vs. $\mu_0H/T$ data are plotted in Fig.~\ref{sctomag}d. The value of $\beta$ obtained from the scaling is $0.4$, which is close to the expected value $1−\gamma = 0.43$. Similar scaling behavior was recently reported in a Kitaev spin liquid candidate H$_3$LiIr$_2$O$_6$ wherein disorder-induced low energy density of states $N(E) \sim E^{-0.5}$ give rise to a power-law behavior in the low-temperature magnetic susceptibility data \cite{PhysRevB.107.014424}. The scaling behavior of magnetization in  the Kitaev magnet H$_3$LiIr$_2$O$_6$ suggests the realization of a random-singlet state. The absence of hysteresis in the $M$ vs. $\mu_0H $ curve at low $T$ also rules out the existence of spin freezing and impurity in this material (see Fig. \ref{MH_high}a). Fig. \ref{MH_high}b depicts the high-field magnetization isotherm data taken at 1.4 K. We tried to reproduce the high-field magnetization data using both modified Brillouin function and power-law as described here \cite{PhysRevB.90.104426}
 \begin{equation}{\label{MHBrillouin}}
 	M(H)=n_\mathrm{imp} \frac{g}{2}\mathrm{tanh}\left[ \frac{g\mu_\mathrm{B}\mu_\mathrm{0}H }{k_\mathrm{B}(T+\theta_\mathrm{imp})}  \right] +A(\mu_\mathrm{0}H)^{\alpha}
 \end{equation}
 where $M(H)$ is in $\mu_\mathrm{B}$ unit, $n_\mathrm{imp}$ is the fraction of orphan spins, $\theta_\mathrm{imp}$ represents antiferromagnetic interaction between  orphan spins. The $g$ value was fixed to 2.2 from ESR measurements and obtained values for  $n_\mathrm{imp}$ and $\theta_\mathrm{imp}$ are 0.34 and 8 K, respectively (see Fig.\ref{MH_high}b). The $\alpha$ value turns out to be 1.74. The power-law dependent high-field magnetization data indicate that there might be enormous low-energy density of states in this disordered antiferromagnet \cite{PhysRevB.90.104426}. 

\subsection*{Specific heat}

\begin{figure*}[!ht]
	\centering
	\includegraphics[height=13cm, width=16cm]{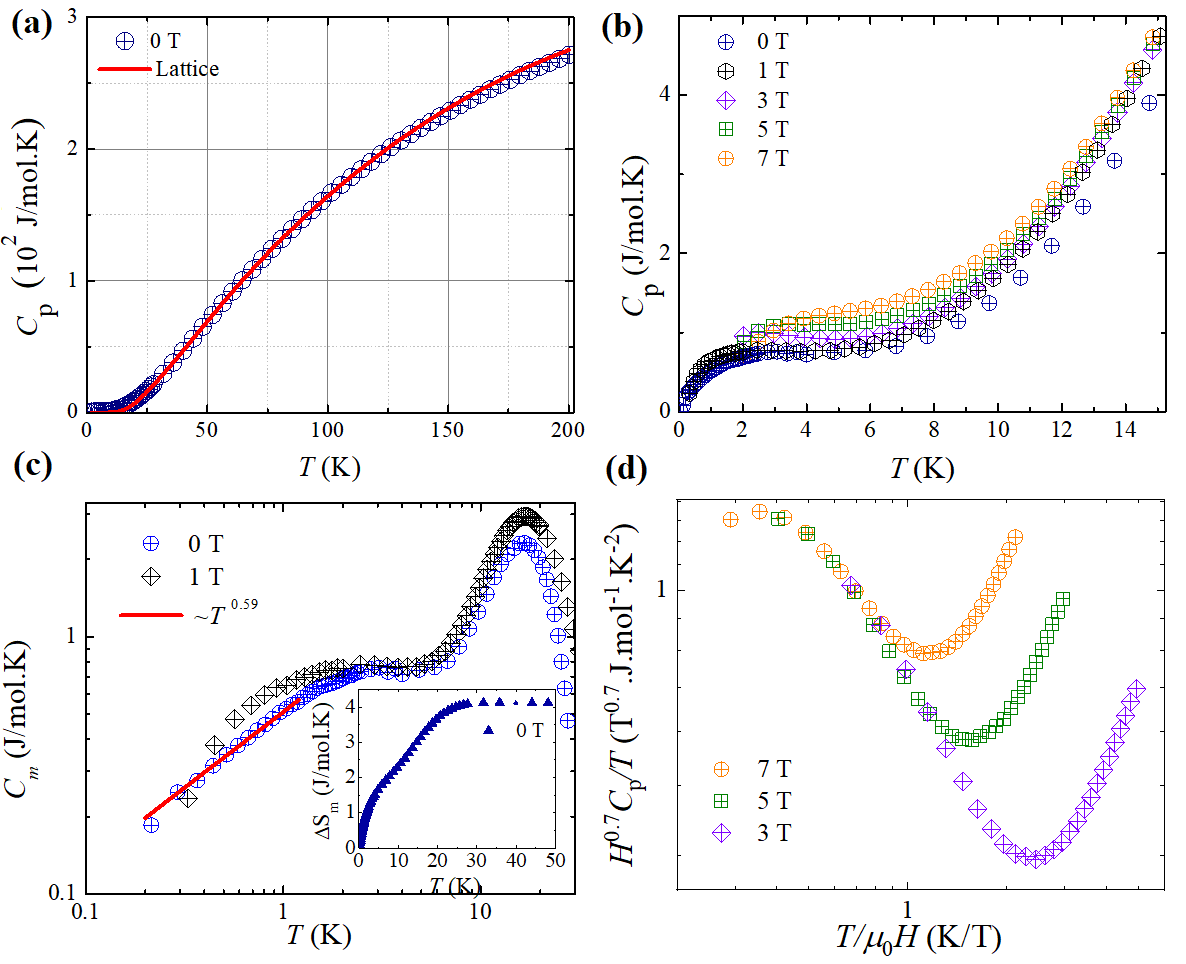}
	\caption{\textbf{(a)} Temperature dependence of specific heat for SCTO in zero field. Red line shows the lattice contribution determined from the combination of Debye and Einstein models as described in the text. \textbf{(b)} Low-temperature specific heat in different applied magnetic fields.  \textbf{(c)} Temperature dependence of magnetic specific heat ($C_\mathrm{m}$) after subtracting the lattice specific heat from total specific heat. Zero-field $C_\mathrm{m}$ follows the power-law $C_\mathrm{m}\sim T^{0.59}$ (red line) at low temperatures. The inset shows entropy release up to 50 K.  \textbf{(d)} The data collapse of specific  heat taken in different magnetic fields onto a single curve suggests the presence of a random-singlet state  that is in agreement with magnetic susceptibility data.}
	\label{specificheatall}
\end{figure*}
\begin{figure*}[!htb]
	\centering
	\includegraphics[height=7.5cm,width=12cm]{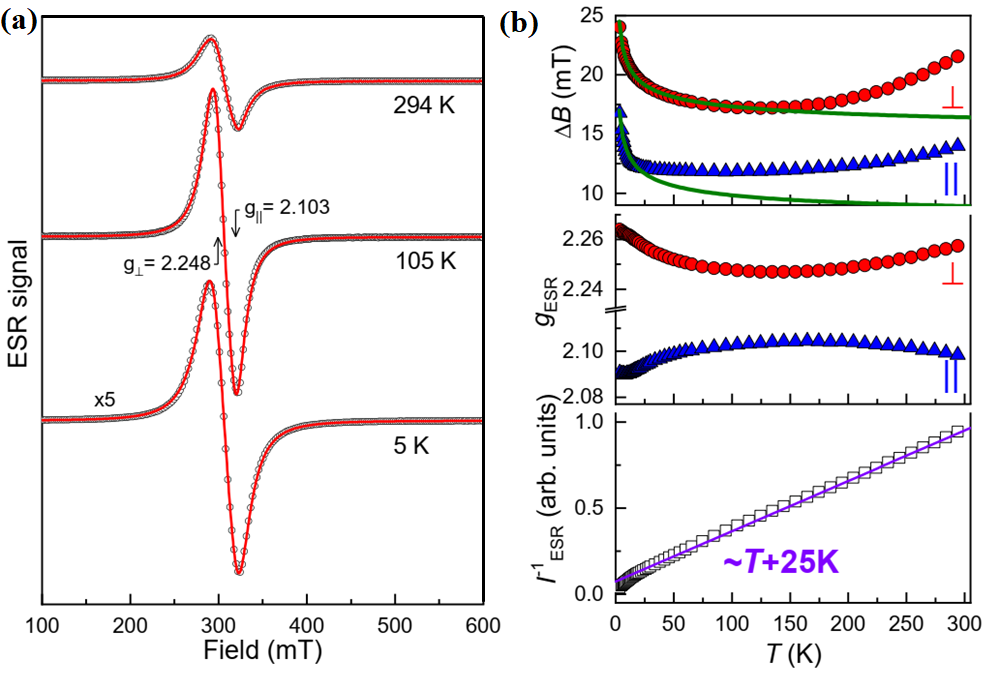}
	\caption{\textbf{(a)} X-band ESR spectra (symbols) at representative temperatures with fit to uniaxial powder-averaged Lorentzians (solid lines). Fitted values of anisotropic $g$-values are indicated for the spectrum at $T = 105$ K. \textbf{(b)} Temperature dependence of fitted parameters linewidth and effective g-value and inverse integrated ESR intensity with Curie-Weiss behavior as indicated by the solid line. Solid lines in the upper panel indicate a pow-law fit $\sim T^{-0.37}$.}
	\label{fig:figesr}
\end{figure*}
Specific heat measurements offer an ideal probe to  track phase transitions and the nature of low-energy excitations at low temperatures in correlated electron materials. Fig. \ref{specificheatall}a and Fig. \ref{specificheatall}b show the temperature dependence of the specific heat in zero field and different applied fields, respectively. The absence of any anomaly down to 64 mK in zero field indicates that there is no long-range magnetic order despite a moderate Curie-Weiss temperature, which suggests that the titled material hosts a quantum disordered state with dominant antiferromagnetic interaction between Cu$^{2+}$ moments in the 3D spin-lattice. At higher temperatures, specific heat is dominated by phonon contribution. In order to extract the phonon contribution, the specific heat data are fitted with a combination of one Debye and three Einstein functions  (solid red line in Fig. \ref{specificheatall}a) as given below \cite{kittel1976introduction,PhysRevB.90.035141}
\begin{equation}
	C_\mathrm{lattice}(T)=C_\mathrm{D}\left[9k_\mathrm{B}\left(\frac{T}{\theta_\mathrm{D}}\right)^{3}\int_{0}^{\theta_\mathrm{D}/T}\frac{x^{4}\mathrm{exp}(x)}{\left[\mathrm{exp}(x)-1\right]^{2}}dx\right]
\end{equation}
\begin{equation*}
	+\sum_{\mathrm{i}} 3RC_{E_\mathrm{i}}\left(\frac{\theta_{E_\mathrm{i}}}{T}\right)^{2}\frac{\mathrm{exp}\left(\frac{\theta_{E_\mathrm{i}}}{T}\right)}{\left[\mathrm{exp}\left(\frac{\theta_{E_\mathrm{i}}}{T}\right)-1\right]^{2}}, 
\end{equation*}
where the first and second term correspond to the Debye and Einstein specific heat, respectively and $\theta_\mathrm{D}$ and $\theta_{E_\mathrm{i}}$ are the Debye and Einstein temperatures, respectively. $C_\mathrm{D}$ and $C_\mathrm{E_i}$ are the Debye and Einstein coefficients, respectively, and $R$ is the universal gas constant. The weightage factor for $C_\mathrm{D}$, $C_\mathrm{E_1}$, $C_\mathrm{E_2}$ and $C_\mathrm{E_3}$ are 1, 3, 5 and 6, respectively. These coefficients are fixed in a way that the ratio of the Debye coefficient and the sum of all Einstein coefficients should $1:n-1$, where $n$ is the number of atoms in a unit cell of SCTO (i.e., 15). The value of $\theta_\mathrm{D}$, $\theta_\mathrm{E_{1}}$, $\theta_\mathrm{E_\mathrm{2}}$ and $\theta_\mathrm{E_\mathrm{3}}$ are $290\pm23$ K, $120\pm1$ K, $279\pm4$ K and $604\pm4$ K, respectively. The lattice contribution is subtracted from the total specific heat data to obtain the magnetic specific heat ($C_\mathrm{m}$). The Fig. \ref{specificheatall}c depicts the low temperature  magnetic specific heat and the inset of  Fig. \ref{specificheatall}c represents the entropy change that is obtained by integrating $C_\mathrm{m}/T$ over the temperature range 0.064 $\leq T\leq$ 50 K for zero field. For a spin-half system, the maximum entropy release should be 5.76 J/mol.K. However, for the present system, the maximum entropy change is $4.13$ J/mol.K in zero field,  which is lower than that expected for $S=1/2$ quantum systems. The residual entropy at low temperatures is attributed to ground state degeneracy and the presence of short range spin correlations that is consistent with appearance of a broad maximum in the magnetic specific heat at low temperatures, which is a common scenario in several disordered quantum magnets \cite{Mustonen2018,PhysRevLett.106.147204,PhysRevLett.125.117206,PhysRevB.95.184425}.

The temperature dependence of specific heat in various magnetic fields provides information concerning spin correlations and magnetic excitations in the ground state of frustrated magnets under study. It is well known that the temperature dependence of specific heat of several quantum spin liquids show a power-law owing to exotic low-energy excitations in the ground state. A random-singlet phase can also give rise to power-law behavior in low-$T$ specific heat data \cite{PhysRevB.101.020406,PhysRevB.103.L241114}. We found that the low-temperature magnetic specific heat data can be reconciled with the power-law $C_\mathrm{m}\sim T^{0.59}$ (see Fig.~\ref{specificheatall}c), suggesting a random-singlet state with low-lying excitations \cite{Khatua2022}. The specific heat $C(H,T)$ shows a unique behavior in random-singlet states and specific heat data in various magnetic fields collapse onto a single scaling curve (for $T/\mu_0H \ll1$)  given by $C(\mu_0H,T)/T \approx \frac{1}{(\mu_0H)^{\gamma}} F(\frac{T}{\mu_0H} )$, where  $F$ is the scaling function \cite{Kimchi2018}. The scaling behavior of the as measured specific heat, $C_\mathrm{p}$,  $(\mu_0H)^{\gamma} C_\mathrm{p}/T $ with $T/\mu_0H$ in SCTO  is demonstrated in Fig. \ref{specificheatall}d, and the obtained value of $\gamma$ is  0.7, which is in line with what is expected in a random-singlet state \cite{Kimchi2018}. The upturn in high  $T/\mu_0H$ values (See Fig.~\ref{specificheatall}d) is due to the phonon contribution.  The observed scaling behavior of the specific heat suggests the realization of a randomness-driven quantum disordered state in SCTO, which is attributed to competing interactions mediated by intrinsic but unavoidable anti-site disorder in the spin-lattice. It is worth mentioning that similar data collapse was also found in a disordered triangular lattice  Y$_2$CuTiO$_6$ \cite{PhysRevLett.125.117206} with a scaling coefficient $\gamma=0.7$. The specific heat in zero field shows a broad maximum around 17 K reflecting the
onset of short-range spin correlation between Cu$^{2+}$ spins at
low temperatures in this  antiferromagnet. Remarkably, the system neither shows the signature of spin-freezing nor magnetic ordering down to 64 mK, despite such a high site dilution, which suggests a quantum disordered ground state in this 3D spin-lattice.

\subsection*{Electron spin resonance}
\begin{figure*}[!htb]
	\centering
	\includegraphics[height=7.5cm,width=18cm]{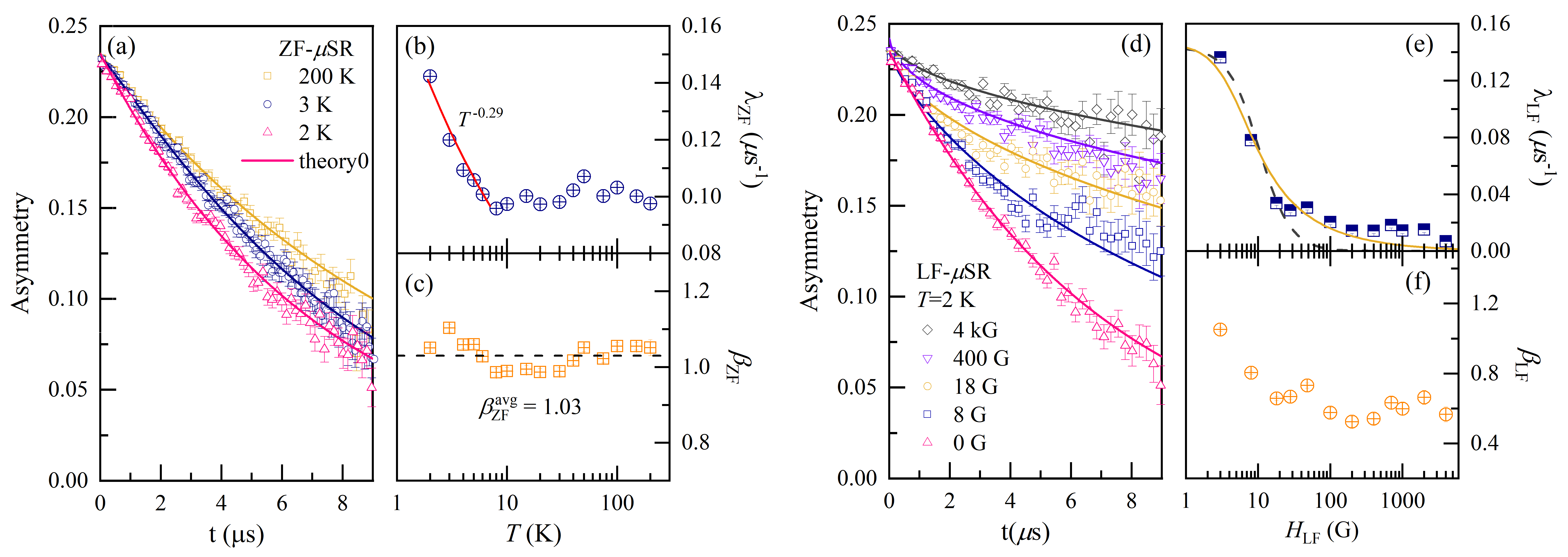}
	\caption{\textbf{(a)} Representative ZF-\musr/ spectra at different temperatures. The solid lines denote the fittings to the data as described in the main text. \textbf{(b)} Temperature dependence of the muon spin relaxation rate $\lambda_\mathrm{ZF}(T)$ in a semi-log scale with a pow-law fit $\sim T^{-0.29}$. \textbf{(c)} Stretched exponent as a function of temperature $\beta_\mathrm{ZF}(T)$. The dashed horizontal line indicates the average value of $\beta_\mathrm{ZF}(T)$.\textbf{(d)} LF-\musr/ spectra at $T=2$ K in various longitudinal fields. The solid lines represent the fits to the data. \textbf{(e)} LF dependence of the muon spin relaxation rate $\lambda_\mathrm{LF}(H_\mathrm{LF})$ in a semilog scale. The solid and dashed curves denote the fittings to $\lambda_\mathrm{LF}(H_\mathrm{LF})$ as described in the main text. \textbf{(f)} Stretched exponent as a function of LF $\beta_\mathrm{LF}(H_\mathrm{LF})$.}
	\label{fig:musrzf}
\end{figure*}
Electron spin resonance (ESR) is a sensitive microscopic probe for magnetism and provides a direct access to the spin dynamics of the Cu$^{2+}$ spins in SCTO. 
The ESR spectra, representing the first derivative of the absorbed microwave power, are  shown in Fig. \ref{fig:figesr}a for a few representative temperatures.
The lines could be well reproduced  using a symmetric Lorentzian shape, averaged for
uniaxial $g$-factor anisotropy (solid red lines). The linewidth $\Delta{B}$ extracted from these spectra is associated with the spin relaxation rate, including spin-spin and spin-lattice interactions. On the other hand, the resonance field
$B_\mathrm{res}$ gives information regarding the $g$-factor and internal fields. Hence, the effective
ESR $g$-factor, given by $g_\mathrm{ESR} = h\nu/\mu_\mathrm{B}B_\mathrm{res}$, is a measure of both the spins’ $g$-factor
and the magnetic molecular-field. For $T = 105$ K, we obtained $g_{\parallel} = 2.103$ and $g_{\perp}= 2.248$, corresponding to an averaged value $g_\mathrm{avg} = \sqrt{(g_{\parallel}^2+ 2g_{\perp}^2 )/3}= 2.201$.
As shown in Fig. \ref{fig:figesr}b,  both $\Delta{B}$ and $g_\mathrm{ESR}$ exhibit a change in their temperature dependence at $T\simeq$ 100 K. The increase of $\Delta{B}$  towards low temperatures indicates the enhancement of Cu$^{2+}$ spin correlations, whereas  spin-lattice relaxation processes dominate the broadening towards high temperatures.
Both perpendicular and parallel component of $\Delta B$ follow power-law behavior at low temperatures with a critical exponent 0.37 (i.e., $\Delta B \sim T^{-0.37}$), indicating a random-singlet ground state \cite{PhysRevB.90.104426}. The ESR intensity $I_\mathrm{ESR}$, which can be measured through the integrated ESR absorption spectra, originates from the static spin-probe susceptibility. Hence, it can be used as an intrinsic microscopic probe of the sample magnetization. As can be seen in the bottom panel of Fig. \ref{fig:figesr}b, the temperature dependence of $I_\mathrm{ESR}^{-1}$ follows a Curie-Weiss formula at high temperatures with the Weiss temperature of $−25$ K, which is consistent with that obtained from magnetic susceptibility data.

\subsection*{Muon spin relaxation}
The magnetization and specific heat results follow a universal scaling behavior which evidences a randomness-driven quantum disordered state in SCTO with quenched disorder. This invokes more sensitive local-probe magnetic investigations that could unambiguously confirm the quantum disordered ground state and enable testing a dynamic scaling behavior. In order to determine the magnetic ground state and spin dynamics, we performed $\mu$SR experiments on SCTO. As displayed in Fig. \ref{fig:musrzf}a, the  ZF-\musr/ spectra show no indication of the presence of static magnetism down to 2 K, such as a coherent muon oscillation signal and 1/3 recovery of the asymmetry at long times. Rather, the asymmetry exhibits an exponential decay in the measurement temperature range. These are consistent with the absence of long-range magnetic ordering down to 2 K, evidenced by the magnetic and thermodynamic properties.
For quantitative analysis, we fit the obtained ZF- and LF-\musr/ spectra with the stretched exponential relaxation function, $P_z (t)=P_z(0)\exp{[-(\lambda_\mathrm{ZF⁄LF} t)^{\beta_\mathrm{ZF⁄LF} }]}$. Here, $\lambda_\mathrm{ZF⁄LF}$ is the muon spin relaxation rate and $\beta_\mathrm{ZF⁄LF}$ is the stretching exponent. 

In Fig. \ref{fig:musrzf}b and Fig. \ref{fig:musrzf}c, we plot the extracted parameters for the ZF-\musr/  results. Above 10 K, the muon spin relaxation rate has a nearly constant value close to $\sim$ 0.10 $\mu$s$^{-1}$, indicating rapidly fluctuating moments. In the paramagnetic state, the spin fluctuation $\nu$ can be evaluated by $\nu=\sqrt{z}JS/\hbar$, where $z=6$ is the nearest-neighbor coordination number and $J$ is the nearest-neighbor exchange interaction. By using $k_\mathrm{B} \theta_\mathrm{CW}=-zS(S+1)J/3$, we obtain the exchange fluctuation rate of $\nu=2.88 \times 10^{12}$ s$^{-1}$. Combined with the relation $\lambda_{ZF}=(2\Delta^2)⁄\nu$, $\lambda_{ZF}$($T>10$ K) $\sim$ 0.10 $\mu$s$^{-1}$ provides the field distribution width of the local magnetic field $\Delta=397$ MHz, less than $\nu=2.88 \times10^{12}$ s$^{-1}$. With decreasing temperature below 10 K, $\lambda_\mathrm{ZF}$ increases gradually down to 2 K. This suggests the slowing down of the Cu$^{2+}$ moment fluctuations by the development of spin correlations below 10 K consistent with thermodynamic results which is generic to QSL candidates \cite{khatua2024spinliquidstateemergent}. Note that the stretched exponent exhibits a nearly temperature independent behavior for all temperatures and has the average value $\beta_\mathrm{ZF⁄LF}^\mathrm{Avg}=1.03$, close to a simple exponential decay. The power-law behavior of $\lambda_\mathrm{ZF}$ ($\sim T^{-0.29}$) at low temperatures indicates a quantum-critical scaling which is consistent with a random-singlet state (see the red line in Fig. \ref{fig:musrzf}b). Similar power-law behavior in zero-field muon relaxation rate was found in the random-singlet material Sr$_2$CuTe$_{1−x}$W$_x$O$_6$ \cite{PhysRevLett.126.037201}.
\begin{figure}[!ht]
	\centering
	\includegraphics[width=8.7cm, height=7.5cm]{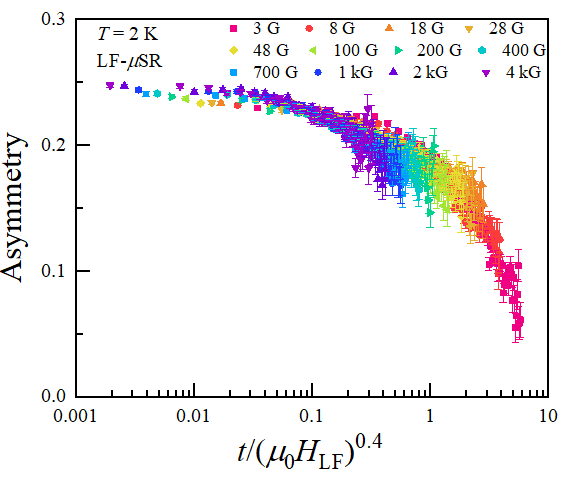}
	\caption{Time-field scaling of LF-\musr/ asymmetries in a semi-log scale at $T=2$ K. The collapse of muon asymmetries onto a single curve demonstrates a randomness-driven dynamic state in \scto/. }
	\label{fig:musrcollapse}
\end{figure}

We turn to the LF-\musr/ results of SCTO. Fig. \ref{fig:musrzf}d shows the representative LF spectra at $T=2$ K in different fields. Though the long-time relaxation of the \musr/ asymmetry begins to recover by applying a weak LF, the applied LF of 4 kG is insufficient to fully decouple the muon spin from internal local fields. The residual relaxation suggests that the muon spin relaxation is mainly dominated by dynamically fluctuating spins, which is consistent with the absence of long-range magnetic order down to 2 K.
In Fig. \ref{fig:musrzf}e and  Fig. \ref{fig:musrzf}f, we summarize the extracted parameters in a semi-log scale. It is well established that, for the simple exponential correlation function, $\lambda_\mathrm{LF}(H_\mathrm{LF})$ can be described by the Redfield formula $\lambda_\mathrm{LF}(H_\mathrm{LF})=2\Delta^2 \nu⁄{(\nu^2+\gamma_\mu^2\mu^2_0H_\mathrm{LF}^2)}$. However,  $\lambda_\mathrm{LF}(H_\mathrm{LF})$ is not reproduced by the Redfield formula (the dashed line in Fig. \ref{fig:musrzf}e). This signifies the presence of an algebraic term in the spin correlation function at low temperatures, $S(t)\sim  (\tau⁄t)^x \exp{(-\nu t)}$,  where $\tau$ and $1/\nu$ the early and late cutoff times and $x$ is the critical exponent \cite{PhysRevLett.92.107204,PhysRevB.64.054403,PhysRevB.84.100401}. 
To describe the field evolution of $\lambda_\mathrm{ZF}$, we employ a general expression that can be derived from both semiclassical and full quantum approaches,
\begin{equation}
	\lambda_\mathrm{LF} (H_\mathrm{LF}) =2\Delta^2 \tau^x \int_0^{\infty}t^{-x} \exp{(-\nu t)} \cos{(2\pi \mu_0 \gamma_{\mu} H_\mathrm{LF} t)}dt.
	\label{eqn:musr}
\end{equation} 

As shown in Fig. \ref{fig:musrzf}e, $\lambda_\mathrm{LF} (H_\mathrm{LF})$ is well reproduced by Eq. ~\ref{eqn:musr} with $x=0.4029$ and $\nu=2.5043 \times 10^6$ Hz (the solid lines in Fig. \ref{fig:musrzf}a). This indicates that the spin auto-correlation function exhibits a power-law-like decaying at low temperatures, which is slower than $S(t)\sim \exp{(-\nu t)}$ with $v=2.88 \times 10^{12}$ Hz at high temperatures. The slowing down of the spin fluctuations reflects the long-time spin correlations at low $T$. Based on these observations, we infer that SCTO has the dynamically fluctuating ground state at least down to 2 K.
We note that the LF dependence of the stretching exponent is somewhat different from the expectation for quantum spin liquid candidates. It is expected that $\beta_\mathrm{LF}$ increases with increasing field because of the quenched inhomogeneous local fields by an LF in disordered  quantum magnets. However, $\beta_\mathrm{LF} (H_\mathrm{LF})$ for SCTO manifest an opposite behavior to the typical behavior. This may be due to exchange randomness by the site disorder between Cu$^{2+}$ and Ta$^{5+}$ ions. The exchange randomness induces an inhomogeneous local field distribution, so that a weak magnetic field can lead to the partial modulation of internal local fields. Therefore, the spatially inhomogeneous local field distributions could be the origin of the LF dependence of $\beta_\mathrm{LF}$.

Figure \ref{fig:musrcollapse} displays the LF asymmetry as a function of $t/(\mu_0H)^\gamma$. As expected for the randomness-induced  disordered state, we observe the time-field scaling behavior of the LF-\musr/ spectra in the plot against $t/(\mu_0H)^\gamma$ with $\gamma=0.4$. The collapse of the LF-$\mu$SR asymmetries into a single curve implies the power-law contribution to the spin correlation function $S(t)\sim  (\tau⁄t)^x \exp{(-\nu t)}$ \cite{PhysRevLett.77.1386,AmitKeren2004}. The scaling exponent $\gamma=0.4$ is comparable to that determined from the isothermal magnetization, which suggests that all the experimental results are consistent taken on complementary scales and point towards  the realization of a random-singlet state in this material. The  collapse  of  LF  $\mu$SR asymmetry curves at low temperatures is a clear signature of cooperative spin dynamics, a characteristic feature of disordered state in correlated materials in the proximity of a quantum critical point, which adds further credence to the thermodynamic results \cite{PhysRevLett.77.1386,AmitKeren2004,yaouanc2011muon}. A random site-dilution of the underlying spin-lattice accounts for the random network of spins. Although this does not give rise to  extra exchange couplings, a random pattern of connected spins can lead to bond randomness and eventually engender random singlets \cite{PhysRevLett.115.147202,PhysRevB.92.134407,Kimchi2018}.  Our $\mu$SR results provide strong signature of a randomness-driven dynamic disordered ground state in this site-diluted 3D antiferromagnet.\\

\section*{IV. SUMMARY}
We have successfully synthesized polycrystalline samples of \scto/ and characterized their magnetic properties down to 64 mK. The Cu$^{2+}$ ($S=1/2$) moments constitute a disordered nearly cubic lattice with sizable antiferromagnetic interaction between constituent spins, and the system does not undergo a phase transition down to 64 mK. Despite substantial site dilution between Cu$^{2+}$ and Ta$^{5+}$ ions, \scto/ shows a dynamic ground state owing to exchange randomness driven by quenched disorder in this  antiferromagnet. This diluted network of spin $1/2$ ions forms a random-singlet phase which is confirmed by data collapse of temperature- and magnetic-field-dependent specific heat data. The power-law behavior of magnetic susceptibility data at lower temperatures also supports the formation of a random-singlet state. The presence of a broad maximum in specific heat and a lower value of entropy at low temperatures compared to that expected for $S=1/2$ systems suggests the persistence of short-range spin correlations, which is supported by the enhancement of the ZF $\mu$SR relaxation rate at low temperatures. The power-law spin auto-correlation function and data collapse of time- and longitudinal-field-dependent muon asymmetry further validates the realization of randomness-induced  quantum disordered state in this nearly 3D quantum magnets with nearest and next nearest exchange interaction between Cu$^{2+}$ spins.  This offers a promising route to explore disorder-driven correlated quantum states in 2D and 3D quantum magnets.

\section*{Acknowledgements}
We thank M. Gomilšek and S. Kundu for insightful discussions. P. K. acknowledges the funding by the Science and Engineering Research Board, and Department of Science
and Technology, India through Research Grants. The work at SKKU is supported by the National Research Foundation (NRF) of Korea (Grant No. 2020R1A5A1016518). We acknowledge support of the HLD at HZDR, member
of the European Magnetic Field Laboratory (EMFL).

\textit{Note added:} During the revision stage of the present work, the authors were aware of a paper \cite{PhysRevB.110.L060403} reporting spin liquid behavior on the triangular lattice phase of the same material synthesized under high temperature conditions. 
\bibliography{references}

\end{document}